# Design of Magnetic Graphene-ribbon for 100 Tera-bit/inch² Information Storage Media

Norio Ota

Graduate School of Pure and Applied Sciences, University of Tsukuba, *1-1-1 Tenoudai Tsukuba-city 305-8571, Japan*

Magnetic graphene-ribbon is a candidate for realizing future ultra high density 100 tera bit/inch² class data storage media. Multiple spin state analysis was done based on the density function theory. A typical model was a super cell [$C_{80}H_7$] which having bare (radical) carbons on one side zigzag edge, whereas mono hydrogenated on another side. Optimizing atomic configuration, self consistent calculation demonstrated that a total energy of the highest spin state is more stable than that of lower one, which came from exchange coupling between carbons. This analysis suggested a capability of designing magnetic data track utilizing such chemically edge modified graphene ribbon. In order to increase areal magnetization density, bilayer and quadri-layer graphene-ribbon model were analyzed. Detailed calculation resulted that also the highest spin state is the most stable one. Multiplying the layer numbers is effective way to realize and enhance strong magnetism.

**Key words**: magnetic recording, graphene, thin film, first principle calculation, density function theory

## 1. Introduction

Current magnetic data storage[1)-2)] has a density around 1 tera-bit/inch² with 10nm length, 25nm width magnetic bit. Ultimate areal density will be $10^4$ tera-bit/inch² utilizing atomic scale materials[3)] as illustrated in Fig.1. There is a missing link between those two density regions. In order to link those two, one promising candidate is a ferromagnetic molecule dot array having a typical areal bit size of 1 nm by 2.5 nm. Recently, carbon based room-temperature ferromagnetic materials are experimentally reported[4)-9)]. They are graphite and graphene like materials. From a theoretical view point, Kusakabe and Maruyama[10)-11)] proposed an asymmetric graphene-ribbon model with two hydrogen modified (dihydrogenated) zigzag edge carbon showing ferromagnetic behavior. Our previous papers[12)-14)] have reported multiple spin state analysis of graphene like molecules modified zigzag edge carbon by several atomic species. In case of bare (radical) carbon, and dihydrogenated carbon edge molecules, the highest spin state exhibits the most stable spin state, which suggest a capability of strong magnetism. Those carbon based material is very attractive for a future ultra high density (over 100 tera-bit/inch²) magnetic recording media. Especially, graphene-ribbon has a nano-meter width and long straight line looks like a recording track. Here, magnetic properties of chemically edge modified asymmetric graphene-ribbon and geometrical optimization are analyzed based on a first principle theory for designing a future ultra high density magnetic storage.

## 2. Designing ferromagnetic graphene ribbon

Ohldag et al.[9)] recently found that proton ion implanted graphite exhibits strong surface magnetization at a room-temperature. In order to explain such remarkable experiment, we proposed an shadow effect model[13)], which means that proton are partially modified zigzag edge carbon by some masking effect resulting strong magnetism. Such a model is simply illustrated in Fig.2 with adding a mask layer. In an industrial view point, such a mask size and position will be controlled very precisely by some proper resist dot array and/or self organized thin film array. Based on such mask model on a graphene-ribbon, typical calculation example is shown in Fig.3. In order to have a discrete magnetic bit on one straight graphene-ribbon, three carbon edges on one side zigzag edge are bare, whereas another side zigzag edge are all mono hydrogenated as marked by red square. Between two magnetic bits, there is a zero magnetization buffer part (blue marked square) with mono hydrogenated edges on both upper and lower zigzag edges. One periodic super-cell for an infinite length analysis is shown in black marked square as [$C_{80}H_7$]. There are several spin states for the same super-cell. This paper studies which multiple spin state has the lowest total energy suggesting stable magnetism. If high spin state is favorable, we can expect strong magnetic behavior and application to magnetic data storage. This paper also extends to bilayer and quadri-layer ribbon model in order to increase an areal magnetization density per bit on a recording track.

## 3. Calculation Method

We have to obtain the (i) spin density map, (ii) total energy per super-cell, and (iii) optimized atom configuration depending on a respective given spin state $S_z$ to clarify magnetism. Density function theory

(DFT)[15),16)] based generalized gradient approximation (GGA-PBEPBE)[17)] was applied utilizing Gaussian03 package[18)] with an atomic orbital 6-31G basis set[19)].

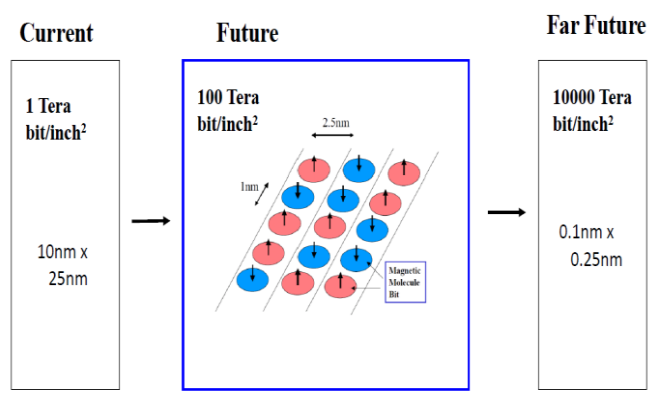

**Fig.1** Future ultra-high density 100 tera-bit/inch² magnetic data storage with bit size of 1 nm by 2.5nm, which may link current and far future density region.

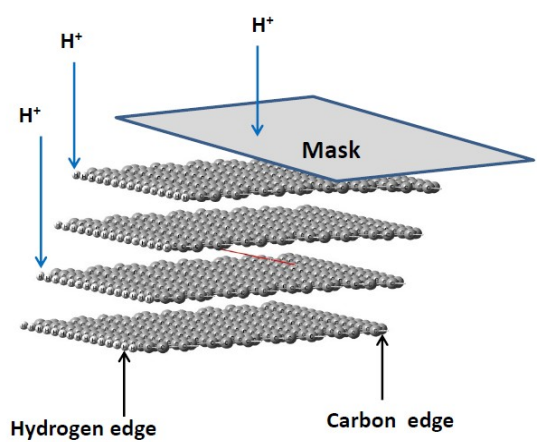

**Fig.2** Proton irradiated graphite model. By a shadow effect using some tiny mask, left side carbon edge is mono-hydrogenated, whereas right side one remains as bare (radical) carbon. Such chemical modification causes strong magnetism.

### 4, Monolayer ribbon

Typical monolayer model is shown in Fig.3. Each super-cell has limited numbers of unpaired electrons, which enable allowable numbers of multiple spin states. In a super cell [$C_{80}H_7$], there are three unpaired electrons which bring two spin state like $S_z=3/2$ and 1/2. These $S_z$ values are installed as spin parameter to start the DFT calculations. Repeating DFT self consistent calculation, minimizing total energy and optimizing atomic configuration, we required a results with relative energy accuracy of at least 10E-8 within 128 calculation cycles. Final obtained spin density maps are shown in Fig.4, comparing (a) $S_z=3/2$ and (b) 1/2, where up-spins are indicated in red (dark gray), and down-spins in blue (light gray) at a 0.001e/A³ contour spin density surface. In case of (a) $S_z=3/2$ (high spin state), we found that up and down spin clouds were very regularly aligned one by one. In contrast, in case of (b) $S_z=1/2$ (low spin state) a very complex spin structure appeared inside a super-cell like up-up and down-down spin pairs. Exchange coupling between up-up (also down-down) brought about a local exchange energy increase, and finally elevated the total graphene-ribbon energy. Exact total super-cell energy of high spin state ($S_z=3/2$) is 13.2kcal/mol lower than that of low spin state ($S_z=1/2$). It should be noted that high spin state is more stable than low spin state, which suggests a possibility of strong magnetism.

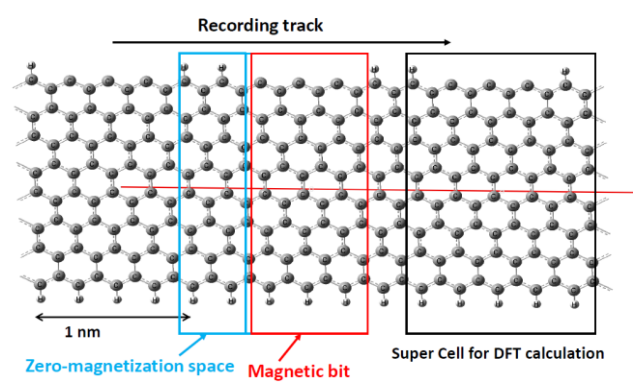

**Fig.3** Design of a magnetic bit on a single layer graphene-ribbon (red marked square). Blue marked area is a buffer to distinct bit to bit. Super-cell for an infinite length ribbon calculation is shown in black mark as [$C_{80}H_7$].

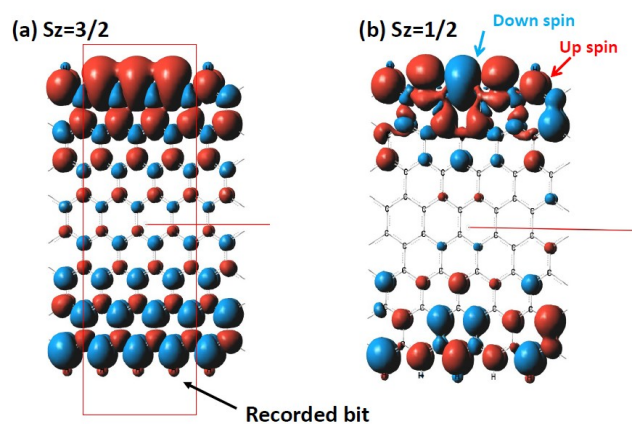

**Fig.4** Spin density map of $S_z=3/2$ in (a) and $S_z=1/2$ in (b) in a super-cell [$C_{80}H_7$], where up-spins are indicated in red (dark gray), and down-spins in blue (light gray). Total energy of high spin state (a) is 13.5kcal/mol lower than that of low spin state (b). High spin state is stable than lower one.

## 5. Multilayer ribbon

In order to increase the areal saturation magnetization per bit, bilayered and quadri layered magnetic graphene-ribbon should be studied. If the highest spin state would be the most stable one, we could expect to increase the areal magnetization very drastically. Here typical bilayer model is shown in Fig.5. Super cell is [$C_{48}H_3$ x 2], which means $C_{48}H_3$ monolayer cell overlaps one by one to construct bilayer. Red squared area show a super-cell, where ball carbon and hydrogen demonstrate a back (second) layer. Zheng et al [20] experimentally observed bilayer graphene zigzag edge structure by means of atomically resolved high-resolution TEM. They indicated that bilayer graphene is stable at a room-temperature with the AA stacking, mostly showing closed (rolled) ends but partially having open zigzag edges. Depending on such observation, our initial atomic configuration was the AA stacking bilayer graphene-ribbon as illustrated in Fig.6 (a), which is a side view with open edges and layer to layer distance of 0.334nm with a typical bulk graphite value. Executing DFT calculation, dynamic forces between atoms gradually enforce two layers moving and sliding each other, and after 95 times self consistent calculation cycles, it finally converged as the AC stacking structure[21] with a distance of 0.304nm as shown in (b). Additionally we could find that bare carbon edges (upper end in (b)) become close together. Comparing with an experimental result[20] after 2000 degree C high temperature annealing, our DFT calculation is limited to quantum force balance at a zero temperature, and may not to have enough force to grow up to the AA stacking.

In our bilayer model, there are three multiple spin state capability as Sz=6/2, 2/2 and 0/2. Result is shown in Fig.7 as spin density map (plane view). In case of the highest spin state (a) Sz=6/2, up and down spin clouds were very regularly aligned one by one. In contrast, in case of (b) Sz=2/2, a very complex spin structure appeared which increases exchange energy and total energy by 1.6kcal/mol per super-cell. The lowest spin state case (Sz=0/2), DFT calculation does not converged, which suggested an unstable spin state. Side view of spin density of Sz=6/2 is shown in Fig.6(c). It should be noted that even in case of bilayer graphene ribbon the highest spin state is more stable, which fortunately suggests a possibility of strong magnetism.

We also checked a quadri-layer case applying [$C_{48}H_3$ x 4] super cell. As resulted in Fig.8, the most stable spin state is the highest spin state Sz=12/2. Unstable one is Sz=0/2. All the bare carbon edge show large up-spin clouds and slightly bending to be close each together. It should be noted that multiplying layer numbers of magnetic graphene-ribbon is an effective method to realize three dimensional molecular size magnetic recording media.

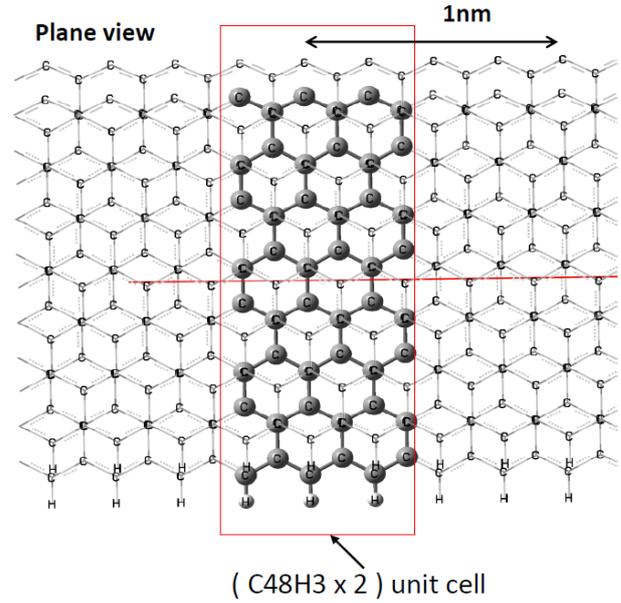

**Fig.5** Plane view of a bilayer magnetic graphene-ribbon with the AC stacked structure having a super-cell [($C_{48}H_3$) x 2]. Red squared area show a super-cell, where ball symbolized carbon and small ball hydrogen demonstrate a back (second) layer.

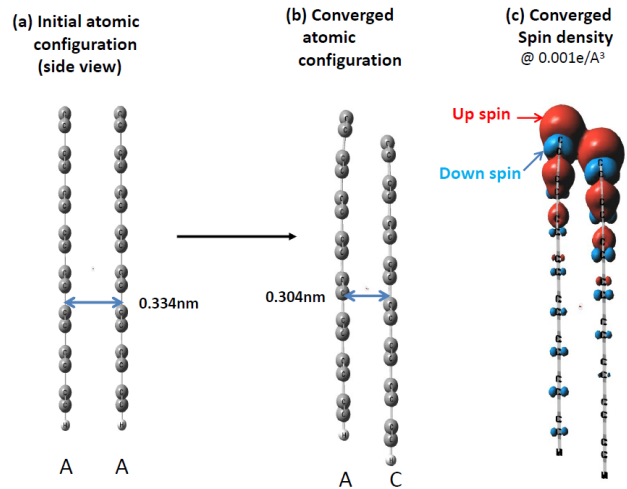

**Fig.6** Side view of bilayer graphene ribbon. Initial atomic configuration is the AA stack parallel structure as shown in (a). Converged calculated result is illustrated in (b) as the AC stack. Also, converged spin density is shown in (c). Up-spin clouds became close together at bare (radical) carbon edges, upper end in (c).

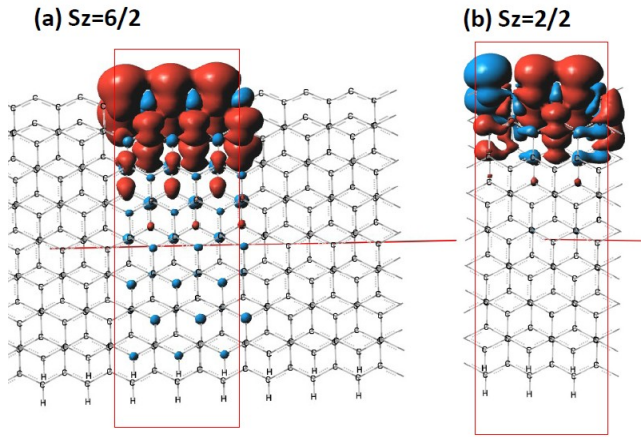

**Fig.7** Spin density map of bilayer graphene-ribbon in a plane view. Total energy of the highest spin state Sz=6/2 in (a) is 1.6 kcal/mol per super-cell lower and stable than that of Sz=2/2 in (b).

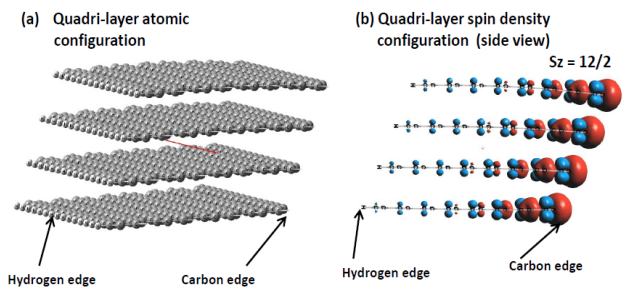

**Fig.8** Quadri-layer graphene-ribbon atomic configuration bird eye view in (a) and spin density side view in (b). The most stable spin state is the highest spin state Sz=12/2. Unstable one is Sz=0/2.

## 6. Conclusion

Carbon based material is very attractive for realizing a future ultra high density 100 tera-bit/inch$^2$ class magnetic information storage media. Especially, graphene-ribbon has a nano-meter width and long straight line looks like a recording track. Here, graphene-ribbon magnetic properties and geometrical optimization are analyzed based on the first principle density function theory. Multiple spin state analysis was done on typical model with bare (radical) carbons on one side zigzag edge, whereas mono hydrogenated on another side. Optimizing atomic configuration, self consistent calculation demonstrated that a total energy of the highest spin state is the lowest one than that of lower spin states, which suggested a possibility of stable and strong magnetism. In order to increase areal magnetization density, bilayer and quadrilayer model were analyzed, which resulted that also the highest spin state is the most stable one. Multiplying the layer numbers is an effective way to enhance and realize strong magnetism.